# Photovoltages and hot electrons in plasmonic nanogaps

Douglas Natelson*,[a,b,c], Charlotte I. Evans[a], Pavlo Zolotavin[d]

[a]Department of Physics and Astronomy, Rice University, 6100 Main St. MS 61. Houston, TX 77005; [b]Department of Electrical and Computer Engineering, Rice University, 6100 Main St. MS 61. Houston, TX 77005; [c]Department of Materials Science and NanoEngineering, Rice University, 6100 Main St. MS 61. Houston, TX 77005; [d]Lam Research Corporation, 4650 Cushing Parkway, Freemont, CA 94538

## ABSTRACT

In metal nanostructures under illumination, multiple different processes can drive current flow, and in an open-circuit configuration some of these processes lead to the production of open-circuit photovoltages. Structures that have plasmonic resonances at the illumination wavelength can have enhanced photovoltage response, due to both increased interactions with the incident radiation field, and processes made possible through the dynamics of the plasmon excitations themselves. Here we review photovoltage response driven by thermoelectric effects in continuous metal nanowires and photovoltage response driven by hot electron production and tunneling. We discuss the prospects for enhancing and quantifying hot electron generation and response via the combination of local plasmonic resonances and propagating surface plasmon polaritons.

## 1. INTRODUCTION

Recent advances in nanofabrication and computational modeling of electrodynamics have led to an enormous growth in plasmonics, the study and application of plasmons, the collective modes of the incompressible electronic fluid in metals and doped semiconductors. Propagating surface plasmon polaritons (SPPs) have proven their use in sensing applications and in manipulating light at lengthscales far below its free-space wavelength. Nanoparticles and structures with nanoscale gaps can host confined plasmon resonances, the excitation of which can produce extremely large local electric fields. The local electric field in proximity to the nanostructure can exceed the amplitude of an exciting incident wave by several orders of magnitude. This enables surface-enhanced spectroscopies, plasmon-based lasing, other quantum optics and optomechanics investigations, and surface photochemical processes. The enhanced photochemical effects can involve "hot" electrons, excitations of the electron gas produced either by conventional absorption or by decay of plasmon excitations where the resulting electrons and holes are, on short (femtosecond) timescales, not thermally equilibrated with the electron gas or the ionic degrees of freedom.

It is of great interest to characterize the physical processes that produce hot electrons, and to delineate between these effects and other processes that can take place in illuminated nanostructures. Time-resolved pump-probe measurements have been of great utility in examining the dynamics of electronic relaxation in metals and plasmonic structures. However, many of the potential applications (e.g., solar fuel production; photodecomposition of targeted species) would involve continuous wave excitation, rather than the high intensity, femtosecond pulses best applied in time-resolved techniques. In the present manuscript, we describe recent investigations involving optoelectronic measurements of plasmonically active nanogap structures that hold promise for improving our understanding optical generation of hot electrons in such systems. We summarize recent results and emphasize that open-circuit photovoltage measurements are advantageous in allowing discrimination against competing effects not directly related to hot electrons.

* natelson@rice.edu; phone 1-713-348-3214; fax 1-713-348-4150; natelson.web.rice.edu

## 1.1 Plasmons in nanowires and nanogaps

Figure 1 shows an example of the nanowire structures employed in these investigations. We begin with a Au nanowire fabricated by electron-beam lithography and e-beam evaporation of 1 nm of Ti and 13-30 nm of Au, depending on the particular experiment. This nanowire serves as a constriction connecting larger electrodes that fan out to wire-bonding pads for electrical measurements. The nanowire's width is around 120-130 nm; measurements and simulations show that this structure has a strong, dipolar transverse plasmon resonance for an incident free-space wavelength of 785 nm (as in our experimental setup). The as-made nanowire has a temperature-dependent electrical resistivity (due to electron-phonon scattering) that allows bolometric measurements of its temperature changes under illumination. The transverse resonance is readily apparent in the polarization dependence of the change in conductivity when illuminated, with the resistance having a form A + B $\cos^2 \theta$, where $\theta$ is the polarization angle defined to be 0 for the incident electric field polarized along the length of the nanowire.

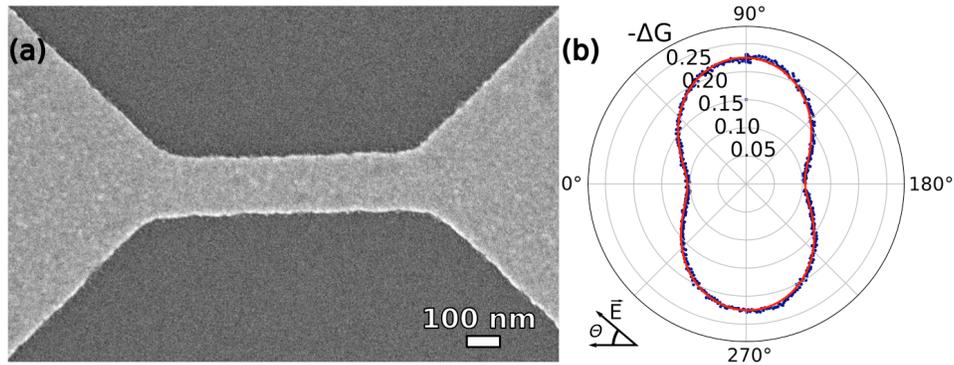

Figure 1. Nanowire constrictions and optical heating. (a) Electron micrograph of a typical nanowire constriction. (b) Change in conductance due to illumination-based heating as a function of incident polarization, with substrate at room temperature. The cosine$^2$ component comes from heating due to the transverse surface plasmon mode of the nanowire.

In the nanogap measurements summarized here, the nanowire is broken via a controlled electromigration process[1] into separate source and drain electrodes, separated by a nanoscale gap intended to be sufficiently small (1-3 nm) to permit a measurable zero-bias interelectrode tunneling conductance. In many prior studies focused on plasmonic properties of nanoscale gaps (as in tip-enhanced Raman spectroscopy[2] or light emission from scanning tunneling microscopy[3]), the dominant mode is an interelectrode dipolar plasmon, as sketched in Fig. 2a. The resonant frequency of such a mode depends critically on the interelectrode distance, and the mode is most readily excited by electric field polarized along the long-axis of the tip, a polarization direction also favored by the "lightning rod" effect[4]. In contrast, in our structures the dominant highly localized modes are found to be multipolar modes localized to the nanogap (Fig. 2b) that are able to be excited by the far field because of hybridization with the dipolar transverse nanowire mode described above[5]. As a result, the hybridized nanogap modes inherit the polarization dependence of the transverse plasmon.

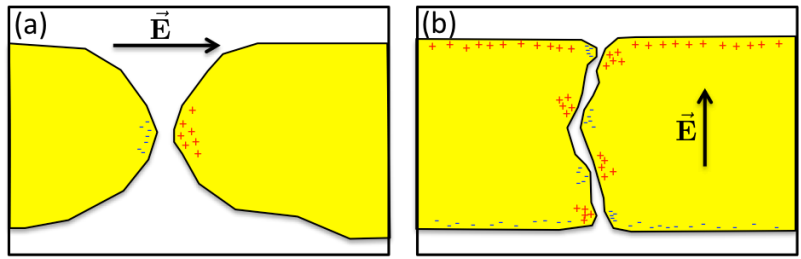

Figure 2. Schematic snapshots of localized plasmon modes in nanogaps. (a) Typical dipolar "tip" mode, relevant in many tip-enhanced spectroscopies, excited when the electric field is polarized as shown. (b) Unusual, highly localized multipolar nanogap modes, excited when polarization is transverse to the nanowire, as shown. Hybridization with the transverse dipolar mode of the nanowire is what allows for excitation of the otherwise optically dark, localized, multipolar modes at the nanogap.[5]

## 1.2 Photo-generated processes in plasmonic structures

The addition of light to a plasmonic structure configured for electronic transport measurements makes several processes possible, as shown schematically in Fig. 3. These can have consequences for the differential conductance, d$I$/d$V$, measured at zero bias or finite bias, the photocurrent $I_{ph}$ if measured in a short-circuit configuration, or other observables. Specific include:

- **Thermal expansion**. Optical absorption can locally increase the temperature of the metal electrodes, and depending on the experimental geometry, this can alter the interelectrode separation, and therefore d$I$/d$V$ at zero bias in the tunneling regime. This cannot produce a zero-bias photocurrent or an open-circuit photovoltage.

- **Photon-assisted tunneling**. The presence of radiation of energy $\hbar\omega$ permits processes that change the electron energy by that amount. Since the tunneling probability between the two electrodes is likely energy dependent, due to both screening/workfunction effects and the energy-dependent density of states of the metals, this can allow tunneling processes that otherwise cannot take place. The Tien-Gordon approach[6] allows analysis of this. The result is a bias-dependent $I_{ph}$ that depends linearly on the incident intensity, and can be nonzero at zero bias if there is an asymmetry in the energy-dependent tunneling between the electrodes on a scale of $\hbar\omega$.

- **Rectification**. In the limit that the tunneling nonlinearity is small on a bias scale of $\hbar\omega/e$, the photon-assisted tunneling picture is equivalent to classical rectification. The incident optical field induces a time-varying voltage at the optical frequency, $V_{opt}$, across the interelectrode gap. For $eV_{opt} \ll \hbar\omega$, the nonlinearity in tunneling, manifested by a nonzero d$^2I$/d$V^2$, leads to a dc photocurrent of magnitude $(1/4)V_{opt}^2$(d$^2I$/d$V^2$). This can be identified by measuring a short-circuit photocurrent $I_{ph}$ as a function of $V_{dc}$ and comparing with the dc dependence of d$^2I$/d$V^2$.[7, 8]

- **Photothermoelectric (PTE) response**. If illumination results in a temperature imbalance across the junction, then through the energy dependence of the tunneling conduction, there can be a net short-circuit $I_{ph}$ given by $-S(T_S-T_D)$(d$I$/d$V$), where $S$ is the net Seebeck coefficient of the junction, and $T_S$ and $T_D$ are the electronic temperatures of the source and drain, respectively. In the open-circuit limit, this produces a photovoltage $-S(T_S-T_D)$.

- **Internal photoemission**. Depending on the energy of the incident photons and the effective work function of the metal electrodes including screening/image charge effects, there can be photoemission. Note that this refers to direct absorption of a photon that creates a photoelectron with appropriate momentum direction to reach the other electrode, regardless of an applied bias. This can lead to a photocurrent in the absence of an applied bias if there is a net imbalance in the photoemission between the two electrodes.

- **Hot electron photocurrent**. This is similar in spirit to both internal photoemission and photothermoelectric response. Unlike the former, the hot electrons may be produced through plasmon decay (and hence can have a different energy distribution than strict photoemitted electrons); unlike the latter, for the hot electrons to remain "hot", they must tunnel before they have had an opportunity to equilibrate with the bulk of the electron gas. The contributing hot electrons produced through plasmonic processes are the ones relevant to the plasmon-enhanced hot electron photochemistry literature. A net short-circuit photocurrent at zero bias requires an imbalance in hot electron production between the two sides of the junction. Analogous to the PTE case, a net short-circuit hot electron current $I_{he}$ would lead to an open circuit photovoltage given by $(I_{pe})/$(d$I$/d$V$), where the differential conductance is at zero bias. If plasmons are a significant contributor to the hot electron generation, then one would expect the photoresponse to share the polarization, wavelength, and laser position dependence of the excitation of plasmons.

We emphasize that disentangling these possible contributions requires careful examination of the dependence of the photoresponse on bias, polarization, position of the incident beam, intensity of the beam, and temperature of the substrate.

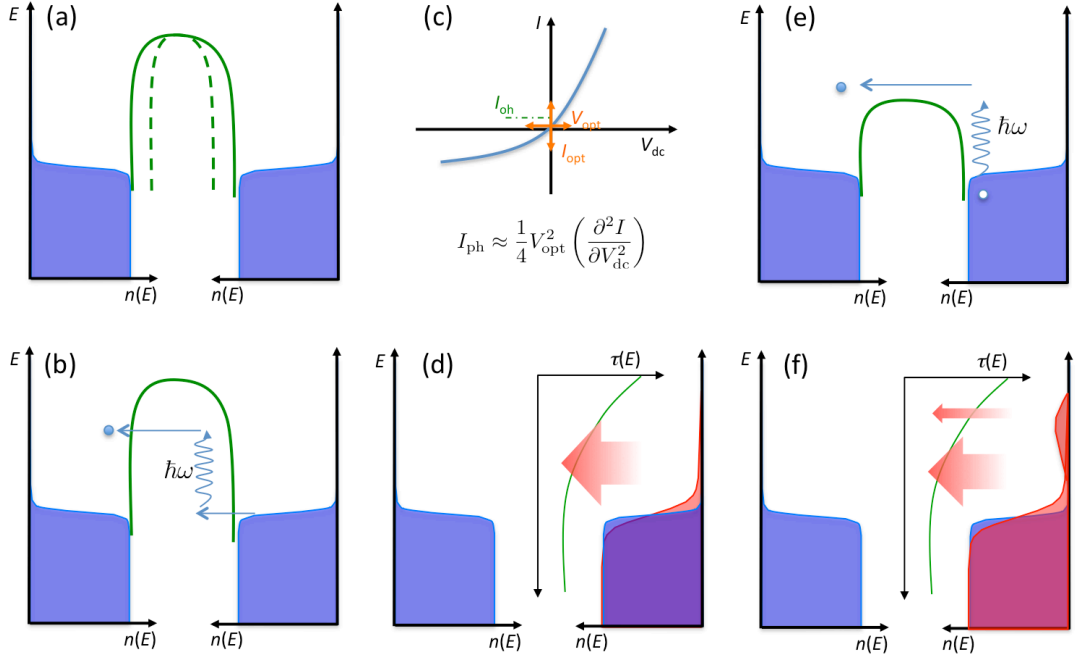

Figure 3. Photo-activated processes in nanoscale junctions. (a) Thermal expansion due to light-induced heating can thin the effective tunnel barrier between source and drain electrodes. (b) Photon-assisted tunneling allows an electron to tunnel into an empty state by acquiring energy from the optical field. (c) Optical rectification can take place in the presence of nonlinear current-voltage characteristics of a nanoscale junction. (d) Photothermoelectric current may be driven if optical heating warms the electronic distribution on one electrode more than the other, with some energy dependent transmission probability setting the magnitude of the effect. (e) Internal photoemission has an electron acquiring sufficient energy from the light field to exceed the effective barrier for tunneling. (f) Hot electron photocurrent relies on the small population of carriers at energies comparable to $\hbar\omega$ that tunnel before thermalizing into the broader electronic distribution.

The expected electronic energy distribution under steady state illumination is important to consider when discerning between different candidate mechanisms. It is well known that under high intensity, pulsed illumination, the electronic distribution can be driven far from thermal equilibrium for tens of femtoseconds[9, 10]. On the tens of fs timescale, electron-electron scattering leads to thermalization of the distribution at some electron temperature, and on ps timescales, energy is transferred to the lattice degrees of freedom. In the steady state, for intensities not in this strongly nonequilibrium regime, the expectation is that the electronic distribution function should be well approximated as a thermal distribution (with a small high energy "tail" out to the photon energy around the Fermi level due to the instantaneous population of electrons and holes that have not yet had time to relax[11]). That distribution is then characterized by a local temperature that is basically the same as the local lattice temperature.

## 2. EXPERIMENTAL PROCEDURES

We consider four device configurations. All of these were fabricated using electron-beam lithography for lateral patterning and e-beam evaporation of 1 nm Ti/15 nm of Au followed by liftoff processing. Substrates were highly doped silicon wafers with oxide layers of either 300 nm or 2 μm of $SiO_2$. Figure 1 shows the basic geometry of the first device type, "short wire", with the transverse width of the nanowire constriction set at 120-130 nm, so that the dipolar transverse plasmon resonance of the wire is excited efficiently by our incident laser (785 nm free space wavelength). The second device type ("short wire nanogap") is created from the short wire by electromigration, breaking the wire into source/drain electrodes separated by a nanoscale gap with the smallest interelectrode separation on the 1-2 nm scale (permitting a measurable tunneling conductance).

For comparison, a third device configuration ("nanowire with gratings") is shown in Fig. 5. The nanowire constriction is the same as in the short wire case, but the larger electrodes are fabricated to have open regions to form a two-slit grating. The purpose of these slits is to break the two-dimensional translational symmetry of the planar electrodes to allow the excitation of propagating surface plasmon polaritons (SPPs). Here Au thickness is 30 nm, to improve SPP propagation. Slit width was chosen to be 300 nm, and slit spacing was ~ 500 nm, a distance that is close to optimal for launch of SPPs. The fourth device configuration ("nanogap with gratings") is the electromigrated version of this.

Devices were wire-bonded to a standard chip carrier and mounted in an optical-access, closed-cycle cryostat (Montana Instruments) for electronic measurement under scanned illumination. The cryostat is integrated into a home-built Raman microscopy system that allows the controlled scanning of a focused (1.8 μm diameter beam spot) 785 nm laser beam across a 10 by 10 micron field of view. Incident power ranged from 10 mW down, regulated by a controlled neutral density filter, and polarization could be rotated via controlled waveplate. Measurements were performed both at room temperature and at cryogenic temperatures.

For most measurements, an optical chopper at 239 Hz was used to modulate the incident intensity. The open-circuit voltage across the device under test was measured using a SR560 voltage preamplifier (100 MΩ input impedance), fed into a lock-in amplifier synced with the chopper. For junctions with measured electrical resistances below the hundreds of MΩ range, the in-phase voltage response far exceeded the out-of-phase voltage response. For the highest tunneling resistances in nanogap devices, however, the lock-in method could produce capacitive phase shifts. Those devices were measured in a true dc configuration (no lock-in amplifier, no chopper). In a large selection of devices, cross-check measurements of short-circuit current under modulated illumination (current preamplifer rather than voltage preamplifier) were performed, and results were found to be consistent with the open-circuit voltage measurements and conductance measurements performed on those same devices.

## 3. RESULTS

### 3.1 Continuous nanowires

Under direct illumination, continuous short nanowires show open-circuit photovoltages due to photothermoelectric effects, as reported elsewhere[12]. The incident laser functions as a scannable heat source, establishing a local electronic and lattice temperature gradient. Thanks to the interplay between nanostructure geometry and energy-dependent electron mean free path, the local Seebeck coefficient, $S$, depends on nanowire geometry. The junction between the nanowire and the larger pad functions as a thermocouple even though the entire structure is made from the same metal film, leading to a net open-circuit voltage when illuminated. The magnitude of the voltage (or equivalently the short circuit current) is consistent with the estimated local laser-induced temperature rise inferred from bolometric measurements of the nanowire conduction[13, 14], and the expected effects of geometry on $S$, as reported previously.

### 3.2 Nanogaps under direct illumination

When the nanowire has been electromigrated to form a tunneling nanogap, direct illumination leads to greatly enhanced photovoltages compared to the continuous nanowire case. As shown in Fig. 4, the very large photovoltages (often on the mV-scale) are very highly localized to the nanogap region, and show the polarization dependence expected for the localized plasmon modes hybridized with the transverse nanowire plasmon. When nanogaps are sufficiently large that the interelectrode conductance is undetectably small, there is no measurable photovoltage. When measured in a short-circuit configuration, nanogap devices that show this large photovoltage have a short-circuit photocurrent consistent with the open-circuit photovoltage and the measured conductance.

The measured open-circuit photovoltages are linear in incident intensity. Because atomic-scale motions of the atoms at the edges of the nanogap can alter the interelectrode tunneling conductance, it is difficult to perform temperature-dependent studies on individual nanogaps without small device configuration changes. Measurements from room temperature down to cryogenic (unilluminated) substrate temperatures show no obvious temperature dependence, however.

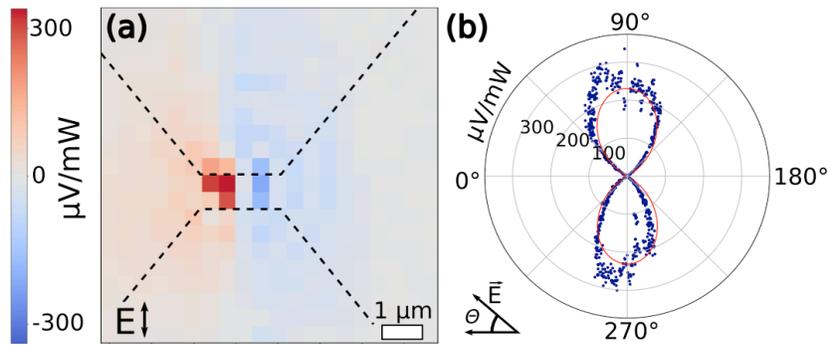

Figure 4. Open-circuit photovoltage of an electromigrated nanogap (initial substrate $T = 5$ K, but quantitatively similar results are obtained at room temperature), per mW of illumination power, as a function of beam position and polarization. Dashed lines are guides to the eye for the approximate optically resolved edges of the metal. The actual nanowire width is ~ 120 nm. The polarization dependence (taken with the beam at one of the red pixels) is consistent with the transverse plasmon mode of the nanowire.

### 3.3 Nanogaps and propagating surface plasmon polaritons

With the motivation of minimizing direct heating of the nanogap region from direct optical absorption of the metal, we have explored delivering plasmon energy to the nanogap via the propagation of SPPs excited remotely using gratings patterned in the electrodes. Using the same bolometric approach as in Figure 1, we could plot the change in junction conductance as a function of laser position and polarization, scanning the junction region as well as the gratings, as shown in Fig. 5.

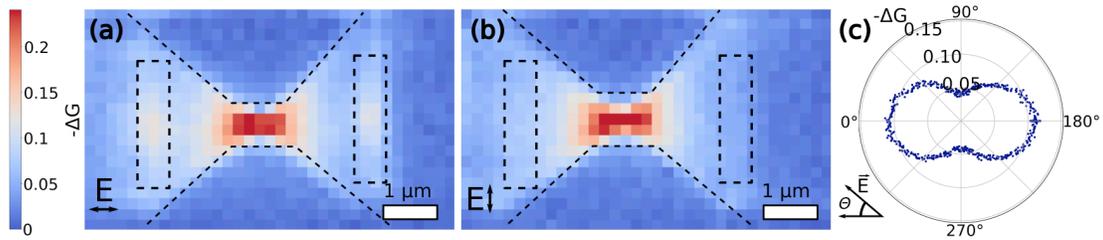

Figure 5. Bolometric heating detected through the device conductance, as a function of laser position and polarization in an unmigrated nanowire (substrate temperature = 40 K, 8 mW illumination power). (a) Polarization oriented to excite SPPs via the gratings, with grating position indicated by the dashed rectangles. (b) Polarization oriented to excite transverse dipolar mode in the nanowire. (c) Polarization dependence of the conductance change when the laser is located at the center of the left-hand grating, showing a clear contribution to junction conductance change when SPPs are excited.

There is a clear conductance change when the laser is positioned over the grating, and that change is maximized when the polarization of the incident beam is aligned in the direction that excites SPPs (perpendicular to the direction that excites the transverse plasmon resonance in the nanowire constriction). Since the total conductance of the structure is largely limited by the nanowire constriction, this indicates that SPPs excited at the gratings are able to propagate to the nanowire region, detectable through their bolometric effect on the nanowire conductance. A detailed study of the magnitude of the conductance change as a function of the distance between the grating and the nanowire center, including computational electrodynamics and thermal transport calculations, has been reported elsewhere.[15] That study demonstrates that, with 2 μm oxide under the devices, the SPPs readily reach the nanowire from gratings more than 10 μm away.

In Fig. 6, we plot a preliminary example open-circuit photovoltage map obtained from such structures following electromigration of the nanowire to form a nanoscale gap, at a substrate temperature of 40 K. With the thicker gold layer required to obtain efficient SPP propagation, it has proven challenging to obtain large interelectrode tunneling resistances reliably. For the device shown, the post-migration interelectrode resistance is approximately 100 kΩ, and the *direct-illumination* open circuit photovoltage is ~ 1 µV/mW pre-migration, ~ 100 µV/mW post-migration.

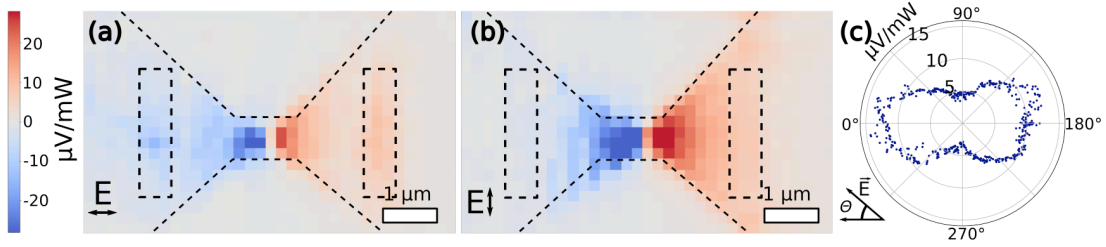

Figure 6. Open-circuit photovoltage in an electromigrated nanogap as a function of laser position and polarization (substrate temperature = 40 K). (a) Polarization oriented to excite SPPs via gratings. (b) Polarization oriented to excite transverse plasmon mode of nanowire. (c) Polarization dependence of open-circuit when laser is located at center of left-hand grating.

As is clear from the plot, illuminating the gratings produces an open-circuit photovoltage, with a polarization dependence such that the voltage magnitude is maximized when the incident polarization is aligned to excite the SPPs. The ratio of pre- and post-migration on-grating voltages is very similar to the ratio of pre- and post-migration direct illumination voltages. Below we discuss candidate mechanisms responsible for the open-circuit voltages in the various cases.

## 4. DISCUSSION AND FUTURE PROSPECTS

Consider the directly illuminated nanogaps first. We now look at possible mechanisms for the generation of this comparatively enormous photovoltage, a thousand times larger than the typical photothermoelectric voltage observed in continuous nanowire structures. The fact that there is an open-circuit voltage rules out any effect of optically-driven thermal expansion or contraction of the interelectrode separation as responsible. The lack of any obvious correlation with differential conductance or higher derivatives of $I$ with $V$ at zero bias, suggests that optical rectification cannot be the mechanism.

Conventional photothermoelectric response seems to be ruled out due to the magnitude of the effect. From the previous bolometric studies of the heating of illuminated nanowire junctions[13, 14], the effective electronic (and lattice) temperature of the metal is tightly constrained. Very large enhancement of the metal $S$ in the tunneling regime would not be expected, and in general, the Seebeck coefficient of tunneling junctions[16, 17] is comparatively small. Contamination with adsorbates or residual titanium oxide from the adhesion layer does not appear quantitatively capable of generating a sufficiently large Seebeck coefficient to produce such voltages, nor does such contamination seem likely to be reproducible in many devices.

Internal photoemission is an intriguing possibility, though seemingly unlikely. This would require the direct generation of photoelectrons with sufficient energy to escape from one electrode, and with momentum directed across the interelectrode gap. Given the normal incidence of the exciting photons and the dependence on the polarization of the incident beam, this seems unlikely.

The most likely explanation, put forward previously[12], is that the open-circuit photovoltage is the result of the tunneling of hot electrons. As discussed by others[11, 18], under steady state illumination, there is a high-energy "hot" carrier population determined by the incident intensity, the absorption at the incident frequency, and the energy-dependent carrier energy relaxation rate. Integrating over the energy-dependent tunneling transmission probability, one finds that for asymmetric illumination and production of hot electrons in the two electrodes, there is a hot electron photocurrent, $I_{ph}$. In an open circuit configuration, the electrodes build up an open circuit photovoltage $V_{ph}$ such that $|V_{ph}| = I_{ph} / (dI/dV$ at $V = 0)$. The polarization dependence to the hot electron open-circuit photovoltage would then originate with the polarization-dependent rate of hot electron generation due to plasmon-enhanced absorption. Reasonable estimates of intensities, absorption, and electronic transmittances give numbers in accord with the measured range of values[12], though detailed quantitative calculations for particular atomic-scale arrangements would be exceedingly challenging.

The open-circuit photovoltage response from the nanogaps excited remotely via grating-launched SPPs is consistent with this picture. While the overall photovoltages in the particular device shown in Fig. 6 are not as large as those in Fig. 4, they remain greatly enhanced compared to those of the unmigrated nanowire. When the gratings are illuminated, optical energy is reaching the nanogap in the form of propagating SPPs. Significant, relatively large (tens of microvolts) photovoltages are seen with a polarization dependence matching that of the grating excitation efficiency, implying that these are generated by the SPPs somehow.

While challenging to model in detail, this configuration is promising for multiple reasons. As intended, the grating approach is a means of getting optical energy to the nanogap while minimally perturbing the steady-state effective electron and lattice temperatures of the metal. The open-circuit voltage obtained in this configuration, combined with modeling, should make it possible to determine the efficiency and mechanism of hot electron generation at the nanogap, since direct absorption is no longer a possible contributor. At issue is whether hot electrons are produced simply by the dissipation of the metal in the propagation of the SPPs, or whether there are more efficient processes (e.g., quantum decay of the SPPs[19, 20]) at work. Similar investigations have been conducted using pulsed lasers and gratings etched on tips[21, 22], but the planar geometry here and the open-circuit configuration avoid possible complications associated with scanned probe microscopies and the other optical effects described in Sect. 1.2. It is hoped that future investigations can reveal much about the creation and dynamics of hot electrons, for their possible uses and to better understand hot electron photochemistry in nanostructures.

## ACKNOWLEDGMENTS

P.Z. and D.N. acknowledge support from ARO Award W911- NF-13-0476 and from the Robert A. Welch Foundation Grant C-1636. C.E. acknowledges support from NSF GRFP DGE1450681. DN also acknowledges support from NSF ECCS-1704625.

## REFERENCES

[1] Park, H., Lim, A. K. L., Alivisatos, A. P. *et al.*, "Fabrication of metallic electrodes with nanometer separation by electromigration," Applied Physics Letters, 75(2), 301-303 (1999).
[2] Hayazawa, N., Saito, Y., and Kawata, S., "Detection and characterization of longitudinal field for tip-enhanced Raman spectroscopy," Applied Physics Letters, 85(25), 6239-6241 (2004).
[3] Pierce, D. T., Davies, A., Stroscio, J. A. *et al.*, "Polarized light emission from the metal-metal STM junction," Applied Physics A, 66(1), S403-S406 (1998).
[4] Liao, P. F., and Wokaun, A., "Lightning rod effect in surface enhanced Raman scattering," The Journal of Chemical Physics, 76(1), 751-752 (1982).
[5] Herzog, J. B., Knight, M. W., Li, Y. *et al.*, "Dark Plasmons in Hot Spot Generation and Polarization in Interelectrode Nanoscale Junctions," Nano Letters, 13(3), 1359-1364 (2013).
[6] Tien, P. K., and Gordon, J. P., "Multiphoton Process Observed in the Interaction of Microwave Fields with the Tunneling between Superconductor Films," Physical Review, 129(2), 647-651 (1963).
[7] Tu, X. W., Lee, J. H., and Ho, W., "Atomic-scale rectification at microwave frequency," The Journal of Chemical Physics, 124(2), 021105 (2006).
[8] Ward, D. R., Huser, F., Pauly, F. *et al.*, "Optical rectification and field enhancement in a plasmonic nanogap," Nature Nanotechnology, 5(10), 732-736 (2010).
[9] Schoenlein, R. W., Lin, W. Z., Fujimoto, J. G. *et al.*, "Femtosecond studies of nonequilibrium electronic processes in metals," Physical Review Letters, 58(16), 1680-1683 (1987).
[10] Qiu, T. Q., and Tien, C. L., "Heat Transfer Mechanisms During Short-Pulse Laser Heating of Metals," Journal of Heat Transfer, 115(4), 835-841 (1993).
[11] Kornbluth, M., Nitzan, A., and Seideman, T., "Light-induced electronic non-equilibrium in plasmonic particles," The Journal of Chemical Physics, 138(17), 174707 (2013).

[12] Zolotavin, P., Evans, C., and Natelson, D., "Photothermoelectric Effects and Large Photovoltages in Plasmonic Au Nanowires with Nanogaps," The Journal of Physical Chemistry Letters, 8(8), 1739-1744 (2017).
[13] Herzog, J. B., Knight, M. W., and Natelson, D., "Thermoplasmonics: Quantifying Plasmonic Heating in Single Nanowires," Nano Letters, 14(2), 499-503 (2014).
[14] Zolotavin, P., Alabastri, A., Nordlander, P. *et al.*, "Plasmonic Heating in Au Nanowires at Low Temperatures: The Role of Thermal Boundary Resistance," ACS Nano, 10(7), 6972-6979 (2016).
[15] Evans, C. I., Zolotavin, P., Alabastri, A. *et al.*, "Quantifying Remote Heating from Propagating Surface Plasmon Polaritons," Nano Letters, 17(9), 5646-5652 (2017).
[16] Smith, A. D., Tinkham, M., and Skocpol, W. J., "New thermoelectric effect in tunnel junctions," Physical Review B, 22(9), 4346-4354 (1980).
[17] Leavens, C. R., and Aers, G. C., "Vacuum tunnelling thermopower: Normal metal electrodes," Solid State Communications, 61(5), 289-295 (1987).
[18] Fung, E. D., Adak, O., Lovat, G. *et al.*, "Too Hot for Photon-Assisted Transport: Hot-Electrons Dominate Conductance Enhancement in Illuminated Single-Molecule Junctions," Nano Letters, 17(2), 1255-1261 (2017).
[19] Manjavacas, A., Liu, J. G., Kulkarni, V. *et al.*, "Plasmon-Induced Hot Carriers in Metallic Nanoparticles," ACS Nano, 8(8), 7630-7638 (2014).
[20] Govorov, A. O., Zhang, H., Demir, H. V. *et al.*, "Photogeneration of hot plasmonic electrons with metal nanocrystals: Quantum description and potential applications," Nano Today, 9(1), 85-101 (2014).
[21] Giugni, A., Torre, B., Toma, A. *et al.*, "Hot-electron nanoscopy using adiabatic compression of surface plasmons," Nature Nanotechnology, 8, 845 (2013).
[22] Müller, M., Kravtsov, V., Paarmann, A. *et al.*, "Nanofocused Plasmon-Driven Sub-10 fs Electron Point Source," ACS Photonics, 3(4), 611-619 (2016).